# Electronic Compensation Technique to Mitigate Nonlinear Phase Noise

Keang-Po Ho, *Member, IEEE*, and Joseph M. Kahn, *Fellow, IEEE*

*Abstract*—Nonlinear phase noise, often called the Gordon-Mollenauer effect, can be compensated electronically by subtracting from the received phase a correction proportional to the received intensity. The optimal scaling factor is derived analytically and found to be approximately equal to half of the ratio of mean nonlinear phase noise and the mean received intensity. Using optimal compensation, the standard deviation of residual phase noise is halved, doubling the transmission distance in systems limited by nonlinear phase noise.

## I. INTRODUCTION

GORDON and Mollenauer [1] showed that when optical amplifiers are used to compensate for fiber loss, the interaction of amplifier noise and the Kerr effect causes phase noise, even in systems using constant-intensity modulation. This nonlinear phase noise, also called the Gordon-Mollenauer effect, corrupts the received phase and limits transmission distance in systems using phase-shift keying (PSK) or differential phase-shift keying (DPSK) [2]-[4]. These classes of constant-intensity modulation techniques have received renewed attention recently for long-haul and/or spectrally efficient WDM applications [5]-[7].

It has been shown recently that the received intensity can be used to compensate the nonlinear phase noise [8]-[9]. Previous compensation methods have used a nonlinear optical component [8] or a phase modulator [9]. In this paper, we describe how to perform the compensation using electronic circuits. We analytically derive the optimal correction factor for this electronic compensation, which can also be applied to optimize the methods of [8]-[9]. We show analytically that optimal compensation can halve the standard deviation (STD) of the nonlinear phase noise, doubling the transmission distance in systems whose dominant impairment is nonlinear phase noise.

The remainder of this paper is organized as follows. Section II introduces the electronic technique for compensation of nonlinear phase noise, and Section III presents a derivation of the optimized compensator and provides numerical results. Sections IV and V present discussion and conclusions, respectively.

## II. ELECTRONIC COMPENSATION OF NONLINEAR PHASE NOISE

We consider a system with multiple fiber spans using an optical amplifier in each span to compensate for fiber loss. For simplicity, we assume that each span is of the same length, and that an identical optical power is launched into each span. Following the model of [1], we neglect the effects of dispersion. In the linear propagation regime, the electric field launched in the $k$th span is equal to $E_k = E_0 + n_1 + n_2 + \ldots + n_k$, $k = 1\ldots N$, where $E_0$ is the transmitted signal, and $n_k$, $k = 1\ldots N$, is the complex amplifier noise at the $k$th span. For a system using binary phase shift-keying (BPSK), $E_0 \in \pm A$. The variance of $n_k$ is $E\{|n_k|^2\} = 2\sigma^2$, $k = 1\ldots N$, where $\sigma^2$ is the noise variance per span per dimension. In the linear regime, ignoring the fiber loss of the last span and the amplifier gain required to compensate it, the signal received after $N$ spans is $E_N = E_0 + n_1 + n_2 + \ldots + n_N$.

Nonlinear phase noise is accumulated span by span, and the overall nonlinear phase shift is equal to [1]

$$\phi_{NL} = \gamma L_{eff} \left\{ |E_0 + n_1|^2 + |E_0 + n_1 + n_2|^2 + \cdots \right.$$
$$\left. \cdots + |E_0 + n_1 + \cdots + n_N|^2 \right\} \quad (1)$$

where $\gamma$ is the nonlinear coefficient of the fiber, and $L_{eff}$ is the effective nonlinear length per fiber span. In the presence of nonlinear phase noise, the received electric field is $E_R = E_N \exp(-j\phi_{NL})$. In PSK systems, an optical phase-locked loop (PLL) [10] can be used to receive the in-phase and quadrature components of the received electric field $E_R$. In DPSK systems, a pair of interferometers [6] can be used to obtain both in-phase and quadrature differential components of the received electric field $E_R$.

Fig. 1a shows an example of a homodyne optical receiver using an optical PLL to detect both in-phase and quadrature components of the received electric field $E_R$ (e.g., see Fig. 5 of [10]). A 90° optical hybrid is used to combine the signal with a phase-locked local oscillator (LO) laser, yielding four combinations with relative phase shifts of 0°, 180°, 90° and 270°. A pair of balanced photodetectors provides in-phase and quadrature photocurrents, $i_I$ and $i_Q$, representative of $\cos(\phi_R)$ and $\sin(\phi_R)$, the corresponding complex components of $E_R$. In systems where the dominant noise source is amplified spontaneous emission from optical amplifiers, a synchronous

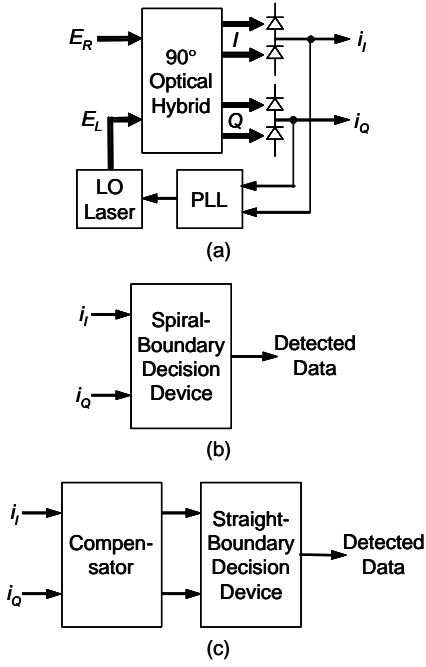

Fig. 1. (a) Typical coherent receiver detecting both in-phase and quadrature components of the received electric field $E_R$. (b) Nonlinear phase noise is compensated by using a spiral-boundary decision device. (b) Nonlinear phase noise is compensated by using a compensator, followed by a straight-boundary decision device.

heterodyne optical receiver can also be used without loss of sensitivity [11].

Figs 2 shows the simulated distribution of the received electric field $E_R$ for a BPSK system with $N = 32$ spans, after detection by a coherent receiver as in Fig. 1a. Note that although the optical PLL of Fig. 1a actually tracks out the mean nonlinear phase shift $<\phi_{NL}>$, nonzero values of $<\phi_{NL}>$ have been preserved in plotting Figs. 2 to better illustrate the nonlinear phase noise. The received optical signal-to-noise ratio (SNR) is $\rho_s = A^2/(2N\sigma^2) = 18$, corresponding to a bit-error rate (BER) of $10^{-9}$ in the linear regime without nonlinear phase noise. In Fig. 2a, the mean nonlinear phase is $<\phi_{NL}> = 1$ rad, corresponding to the maximum mean nonlinear phase shift estimated in [1]. Fig. 2b illustrates the case $<\phi_{NL}> = 2$ rad. The helical-shaped distributions in Figs. 2 arise because the nonlinear phase rotation is correlated with the received intensity [8]-[9]. Figs. 2 also show spiral curves that separate the plane into two decision regions. These decision regions resemble the Yin-Yang logo of Chinese mysticism, and are called the "Yin-Yang detector" below. The Yin-Yang detector uses strictly electronic techniques to compensate nonlinear phase noise, and hence, it differs significantly from the optical [8] and electro-optical [9] compensation techniques considered previously. The optimal spiral curves are derived in the next section.

In this paper, we describe two methods to electronically compensate nonlinear phase noise in a coherent receiver such as in Fig. 1a. The simplest method, shown in Fig. 1b, is the Yin-Yang detector with decision regions separated by spiral curves, like those shown in Fig. 2. Once the mean nonlinear phase shift $<\phi_{NL}>$ is known, one can implement the spiral-boundary decision device using a lookup table. An alternate method, shown in Fig. 1c, employs a compensator that subtracts from the received phase a correction proportional to the received intensity. The compensator is followed by a straight-boundary decision device.

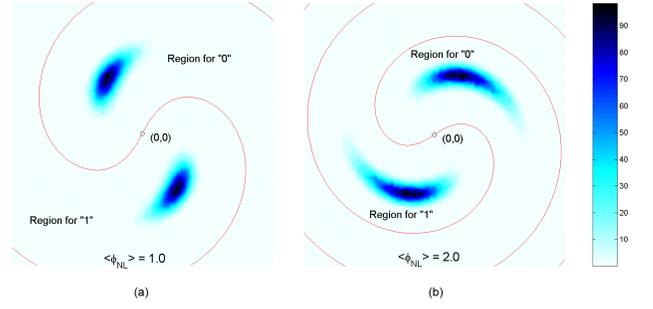

Fig. 2. Simulated distribution and decision regions of received signal with nonlinear phase noise for various mean nonlinear phase shifts: (a) $<\phi_{NL}> = 1$ rad and (b) $<\phi_{NL}> = 2$ rad.

## III. THE OPTIMAL COMPENSATOR

In this section, we derive the optimal compensator for the receiver of Fig. 1c. We then determine the optimized spiral decision boundaries for the receiver of Fig. 1b. The reduction in the STD of nonlinear phase noise is also presented in this section.

In an $N$-span system, to first order, the optimal compensator can be derived by finding a scale factor $\alpha$ to minimize the variance of the residual nonlinear phase shift $\phi_{NL} + \alpha P_N$. The corrected phase estimate is $\phi_R - \alpha P_N$, where $\phi_R$ is the phase of the received electric field $E_R$.

First, we consider a simple mathematical problem. For a real variable $A = |E_0|$ and two complex circular Gaussian random variables $\xi_1$ and $\xi_2$, both $|A + \xi_1|^2$ and $|A + \xi_1 + \xi_2|^2$ are noncentral $\chi$-squared distributed random variables with two degrees of freedom. From [12], the mean and variance of $|A + \xi_1|^2$ are

$$m_{|A+\xi_1|^2} = A^2 + 2\sigma_1^2, \qquad (2)$$

$$\sigma^2_{|A+\xi_1|^2} = f(\sigma_1^2) = 4A^2\sigma_1^2 + 4\sigma_1^4, \qquad (3)$$

where $E\{|\xi_1|^2\} = 2\sigma_1^2$ is the variance of $\xi_1$. In (3), the variance of the random variable $|A + \xi_1|^2$ is defined as a function of $f(\sigma_1^2)$. After some algebra, the covariance between $|A + \xi_1|^2$ and $|A + \xi_1 + \xi_2|^2$ is found to be

$$E\{(|A+\xi_1|^2 - m_{|A+\xi_1|^2})(|A+\xi_1+\xi_2|^2 - m_{|A+\xi_1+\xi_2|^2})\} = f(\sigma_1^2). \qquad (4)$$

The covariance relationship (4) is obvious, because $|A + \xi_1|^2$ does not depend on, and is not correlated with, the random variable $\xi_2$.

Using the above expressions (2)-(4), the variance of the nonlinear phase of (1) is found to be

$$\sigma^2_{\phi_{NL}}(N) = (\gamma L_{eff})^2 \left[ \sum_{k=1}^{N} f(k\sigma^2) + 2\sum_{k=1}^{N}(N-k)f(k\sigma^2) \right]. \quad (5)$$

The first summation of (5) corresponds to all the terms $\sigma^2_{|E_0+n_1+\cdots+n_k|^2}$, the variances of $|E_0+n_1+\cdots+n_k|^2$. The second summation of (5) corresponds to all the covariance terms between $|E_0+n_1+\cdots+n_k|^2$ and

$$|E_0+n_1+\cdots+n_{k+1}|^2+\cdots+|E_0+n_1+\cdots+n_N|^2.$$

Substituting (3) into (5), we obtain

$$\sigma^2_{\phi_{NL}} = \tfrac{2}{3}N(N+1)(\gamma L_{eff}\sigma)^2 \left[(2N+1)|E_0|^2 + (N^2+N+1)\sigma^2\right] \quad (6)$$

Similar to (5), the variance of the residual nonlinear phase shift $\phi_{NL} + \alpha P_N$, is found using (3) and (5) to be

$$\sigma^2_{\phi_{NL}+\alpha P_N}(\alpha) = (\gamma L_{eff})^2 \left[ \sigma^2_{\phi_{NL}}(N-1) + (\alpha-1)^2 f(N\sigma^2) - 2(\alpha-1)\sum_{k=1}^{N-1} f(k\sigma^2) \right]. \quad (7)$$

The optimal scale factor can be found by solving $d\sigma^2_{\phi_{NL}+\alpha P_N}(\alpha)/d\alpha = 0$ to obtain

$$\alpha = 1 + \sum_{k=1}^{N-1} f(k\sigma^2) \Big/ f(N\sigma^2). \quad (8)$$

After some algebra, the optimal scale factor is found to be

$$\alpha = -\gamma L_{eff}\frac{N+1}{2}\cdot\frac{|E_0|^2+(2N+1)\sigma^2/3}{|E_0|^2+N\sigma^2} \approx -\gamma L_{eff}\frac{N+1}{2}. \quad (9)$$

The approximate equality in (9) is valid at high SNR. The variance of the residual nonlinear phase shift is reduced from (6) to

$$\sigma^2_{\phi_{NL}+\alpha P_N} = (N-1)N(N+1)(\gamma L_{eff}\sigma)^2 \times \frac{|E_0|^4+2N\sigma^2|E_0|^2+(2N^2+1)\sigma^4/3}{3(|E_0|^2+N\sigma^2)}. \quad (10)$$

For high SNR and large number of fiber spans, the variance of the residual nonlinear phase shift is

$$\sigma^2_{\phi_{NL}+\alpha P_N} \approx \frac{N^3(\gamma L_{eff}\sigma|E_0|)^2}{3} = \frac{<\phi_{NL}>^2}{6\rho_s} \quad (11)$$

and the variance of the nonlinear phase shift is

$$\sigma^2_{\phi_{NL}} \approx \frac{4N^3(\gamma L_{eff}\sigma|E_0|)^2}{3} = \frac{2<\phi_{NL}>^2}{3\rho_s}, \quad (12)$$

where the mean nonlinear phase shift is

$$<\phi_{NL}> = N\gamma L_{eff}[|E_0|^2 + (N+1)\sigma^2] \approx N\gamma L_{eff}|E_0|^2. \quad (13)$$

The approximation in (12) is the same as that in [1].

Note that the optimal scale factor (9) is approximately equal to one-half the ratio of the mean nonlinear phase shift (13) to the mean received intensity $|E_0|^2 + 2N\sigma^2 \approx |E_0|^2$. From (11) and (12), the phase noise variance can be reduced by a factor of four by using optimal compensation of the nonlinear phase noise.

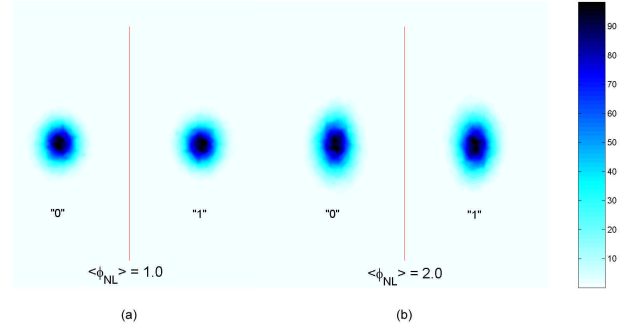

Fig. 3. Simulated distribution of corrected signal using optimal correction factor for various mean nonlinear phase shifts: (a) $<\phi_{NL}> = 1$ rad and (b) $<\phi_{NL}> = 2$ rad.

To our knowledge, the optimized scale factor (9) and variance of residual phase noise (10) have been derived here for the first time. While the theory of [4] considered a more complicated system with a particular pulse shape, the simple approximation in (6) or (12) may yield more useful insight. We should note that in [9], simulation was used to optimize the scale factor, yielding a result corresponding to (9). Our analytical optimization of the scale factor serves as an independent verification of those simulation results.

Figs. 3 show the distribution of the corrected signal $E_c = E_R\exp(-j\alpha P_N)$, assuming the same parameters as Figs. 2. The distributions shown in Figs. 3 have been rotated by the mean phase $<\phi_{NL} + \alpha P_N>$, so that the decision regions become the right and left half-planes. Comparing Figs. 2 to Figs. 3, we see that the phase correction has dramatically reduced the STD of the nonlinear phase shift. Note that, ignoring a rotation, the phase distribution in Fig. 3b is similar to that in Fig. 2a.

The above derivation yielded the optimal value of the scale factor $\alpha$ for the compensator of Fig. 1c. In the receiver shown in Fig. 1b, the optimized spiral decision boundaries are given by rotated versions of $\phi + \alpha\rho^2 = 0$, where $\rho$ and $\phi$ are the radius and phase in polar coordinates. These optimized decision boundaries are shown in Figs. 2. Besides the number of fiber spans, the parameters that determine the decision boundaries are the mean nonlinear phase shift $<\phi_{NL}>$ and the SNR $\rho_s$.

Decoding the corrected electric field $E_c$ using the half-plane decision regions shown in Figs. 3 is equivalent to decoding the received electric field $E_R$ using the Yin-Yang decision regions shown in Figs. 2.

Fig. 4 shows the STDs $\sigma_{\phi_{NL}}$ and $\sigma_{\phi_{NL}+\alpha P_N}$, given by (6) and (10), as functions of the mean nonlinear phase shift $<\phi_{NL}>$ of (13), for $N = 8$ and 32 spans. Fig. 4 assumes SNR = 18, like Figs. 2 and 3. Fig. 4 also shows the approximations

$$\sigma_{\phi_{NL}} \approx 0.1925 <\phi_{NL}> \text{ and } \sigma_{\phi_{NL}+\alpha P_N} \approx 0.0962 <\phi_{NL}>, \quad (14)$$

obtained from (12) and (11), as dotted lines. When the correction factor (9) is employed, the STD of the residual nonlinear phase shift $\sigma_{\phi_{NL}+\alpha P_N}$ is nearly independent of the number of fiber spans, and is very close to the approximation in (14). For a given value of the mean nonlinear phase shift $\langle\phi_{NL}\rangle$, the STD of the nonlinear phase shift $\sigma_{\phi_{NL}}$ decreases with increasing $N$. For $N = 32$, $\sigma_{\phi_{NL}}$ is indistinguishable from the approximation given by (14). Fig. 4 demonstrates that for large $N$, our phase correction scheme reduces the STD of nonlinear phase noise by a factor of two.

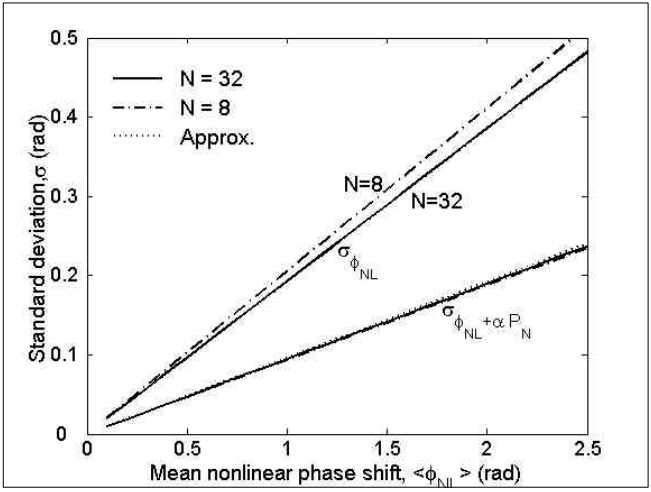

Fig. 4. The standard deviation of nonlinear phase noise (with and without optimal compensation) as a function of the absolute mean nonlinear phase noise $|\langle\phi_{NL}\rangle|$ for $N = 8$ and 32 spans.

## IV. DISCUSSION

Gordon and Mollenauer [1] estimated that the nonlinear phase noise-limited transmission distance is limited to a value such that the mean nonlinear phase shift is $\langle\phi_{NL}\rangle = 1$ rad. From Fig. 4 and (14), this corresponds to a STD of $\sigma_{\phi_{NL}} \approx 0.1925$ rad. Because a mean phase contains neither information nor noise, while the STD of phase $\sigma_{\phi_{NL}}$ is an indicator of system impairment, we can restate the condition for maximum transmission distance in terms of the STD of phase as $\sigma_{\phi_{NL}} \approx 0.1925$ rad. Using our phase correction scheme (or the Yin-Yang detector) and allowing the STD of corrected phase to take on the same value, i.e., $\sigma_{\phi_{NL}+\alpha P_N} = 0.1925$ rad, corresponds to a mean nonlinear phase shift of $\langle\phi_{NL}\rangle = 2$ rad. Because the mean nonlinear phase shift is proportional to the number of fiber spans, as shown in (13), doubling the mean nonlinear phase shift doubles the number of fiber spans, and thus doubles the transmission distance, assuming that nonlinear phase noise is the primary limitation.

While the foregoing discussion has focused on BPSK, the use of DPSK has generated much more interest recently [4]-[9]. In a DPSK system, information is encoded in phase differences between successive symbols, and is decoded using the differential phase $\phi_R(t+T) - \phi_R(t)$, where $T$ is the symbol interval. When the differential phase is corrupted by the nonlinear phase shift difference $\phi_{NL}(t+T) - \phi_{NL}(t)$, the impact of nonlinear phase noise can be compensated by decoding $\phi_R(t+T) - \phi_R(t) - \alpha[P_N(t+T) - P_N(t)]$, where $P_N(t+T) - P_N(t)$ is the power difference between successive symbols. The optimal scale factor for DPSK systems is precisely analogous to that for BPSK systems, and optimal compensation also approximately doubles the transmission distance. The nonlinear phase noise can also be compensated by either a spiral-boundary decision device (as in Fig. 1b) or a compensator followed by a straight-boundary decision device (as in Fig. 1c). The optimal compensator can also reduce the STD of nonlinear phase noise by a factor of two.

A practical coherent receiver, such as the one shown in Fig. 1a, may yield the in-phase and quadrature components $\cos(\phi_R)$ and $\sin(\phi_R)$ [10]. In order to correct the received phase as in $\phi_R - \alpha P_N$, the corrected quadrature components can be calculated using electronic signal processing techniques, as $\cos(\phi_R - \alpha P_N) = \sin(\phi_R)\sin(\alpha P_N) + \cos(\phi_R)\cos(\alpha P_N)$ and $\sin(\phi_R - \alpha P_N) = \sin(\phi_R)\cos(\alpha P_N) - \cos(\phi_R)\sin(\alpha P_N)$.

Other types of nonlinear phenomena may also limit the transmission distance in WDM systems. The interaction of the Kerr effect and optical amplifier noise also induces intensity noise [13], which we have ignored in this paper. Like [1], [8]-[9], this paper ignores all dispersion and filtering effects.

The SNR defined in this paper ($\rho_s$) is defined over a filter bandwidth matched to the signal bandwidth. If the optical SNR (OSNR) is measured using an optical spectrum analyzer over an optical bandwidth $BW_{opt}$, the SNR defined here is related to the measured OSNR as $\rho_s = 2\,OSNR \times BW_{opt}/R_{sym}$, where $R_{sym}$ is the signal symbol rate and the factor of 2 assumes a polarization-insensitive optical spectrum analyzer.

This paper cannot resolve the question whether nonlinear phase noise is the primary limitation for PSK and DPSK systems. Nonetheless, as the electronic compensation technique described here can reduce the impact of nonlinear phase noise, this phase noise becomes less likely to be the dominant impairment.

## V. CONCLUSIONS

In systems using BPSK or DPSK, the impact of nonlinear phase noise can be reduced by using electronic circuits to implement the Yin-Yang decision regions shown in Figs. 2. Equivalently, the received phase can be compensated as described above, in which case the receiver should employ the half-plane decision regions shown in Figs. 3. The optimal compensation factor has been derived analytically for the first time. This compensation halves the STD of the residual nonlinear phase shift, permitting a doubling of the number of fiber spans and of the transmission distance, assuming that nonlinear phase noise is the dominant system impairment.

## ACKNOWLEDGMENTS

We would like to thank the anonymous reviewers of an earlier version of this manuscript for making us aware of references [4], [8], and [9].